# Resonance –Free Light Recycling

You-Chia Chang[1], Samantha P. Roberts[1], Brian Stern[1,2] and Michal Lipson[1*]

[1]Department of Electrical Engineering, Columbia University, New York, New York 10027

[2]School of Electrical and Computer Engineering, Cornell University, Ithaca, New York 14853

* Corresponding author. Electronic mail: ml3745@columbia.edu

**The inability to efficiently tune the optical properties of waveguiding structures has been one of the major hurdles for the future scalability of integrated photonic systems. In silicon photonics, although dynamic tuning has been achieved with various mechanisms[1-10], even the most effective thermo-optic effect[5] offers a refractive index change of only $1.86 \times 10^{-4}$ K$^{-1}$. To enhance this small change, light recycling based on resonators has been employed in order to realize efficient modulators, phase shifters, and optical switches[1,7-11]. However, the resonant enhancement comes at a great cost of optical bandwidth, fabrication tolerance and system scalability. Here we demonstrate a scalable light recycling approach based on spatial-mode multiplexing. Our approach offers a fabrication tolerance of ± 15 nm, in stark contrast to the non-scalable subnanometer tolerance in typical silicon resonators. We experimentally demonstrate light recycling up to 7 passes with an optical bandwidth greater than 100 nm. We realize power-efficient thermo-optic phase shifters that require only 1.7 mW per π, representing more than an 8-fold reduction in the power consumption.**

The scalability of active silicon photonic circuits depends on two major tuning mechanisms of silicon, the thermo-optic effect[2-5] and the plasma dispersion effect[1,6-10], which offer only $\Delta n = 1.86 \times 10^{-4}$ K$^{-1}$ and $-[8.8\times10^{-22}\times n_e + 8.5\times10^{-18}\times n_h^{0.8}]$, respectively ($n_e$ and $n_h$ are the electron and the hole density in cm$^{-3}$). Systems based on active silicon devices that require strong phase and amplitude tunability can be extremely power hungry and hence are constrained in scalability. For example, a chip-scale optical phased array[12-14] with a thousand state-of-art thermo-optic phase shifters[2,3] would consume a prohibitive power of at least tens of watts.

Light recycling using resonators achieves enhanced tuning by allowing light to circulate multiple round trips within the active devices[1,7-11]; however, it requires additional active compensation due to its inherently small optical bandwidth and tight fabrication tolerance. A typical silicon microring modulator with a quality factor of 15,000 has an optical bandwidth of only $\approx$ 0.1 nm[1,7-9]. The reduction of the optical bandwidth in resonators is fundamental, a consequence of requiring constructive interference between the circulating waves within the structure. In addition, because the resonance wavelength of a silicon microring typically shifts by $\approx$ 1 nm per 1 nm of dimensional variation[15], the subnanometer optical bandwidth directly translates to an almost unachievable subnanometer fabrication tolerance[15,16]. Silicon resonators therefore require active compensation to align the resonance frequencies of every individual resonator to the laser frequency. This tuning requirement consumes significant power, limiting the advantage of the resonator-based light recycling.

We show here a broadband light recycling approach that can be used to reduce the power consumption of a photonic integrated circuit without requiring active compensation. Our approach relies on spatial mode multiplexing for enabling multiple light recycling events. In our approach, interference and hence bandwidth reduction are avoided by ensuring that conversion to a different orthogonal spatial mode follows every recycling event. In contrast, in resonator-based approaches, interference always accompanies recycling. In Fig. 1a we show the recycling structure consisting of a multimode bus waveguide integrated with multiple passive mode converters. Figure. 1b shows a schematic of the light path. Light is input into the recycling structure in the $TE_0$ mode, and upon exiting the bus waveguide, it is converted to the $TE_1$ mode via a mode converter and sent back to the bus waveguide in the opposite direction. Upon exiting the bus waveguide, it is then converted to the $TE_2$ mode and sent back to the bus waveguide in the forward direction, and so on. In the recycling structure shown in Fig. 1a, light circulates back and forth a total of 7 times. We design the structure so that light exiting the recycling structure is in the desired $TE_0$ mode. This is done by ensuring that light propagating in the highest order mode is converted back to the fundamental mode. We design the mode converters to pick only one particular mode, raise the mode order by one, and reverse the propagation direction, while keeping all other lower-order modes unperturbed, as shown by the example of the $TE_2$-to-$TE_3$ converter in Fig. 1c. In this example, one of the directional couplers selectively couples the $TE_2$ mode of the bus waveguide to the $TE_0$ mode of the narrow access waveguide[17-22]. This access waveguide is connected to another directional coupler, which selectively couples the $TE_0$ mode of the access waveguide back to the $TE_3$ mode of the bus waveguide.

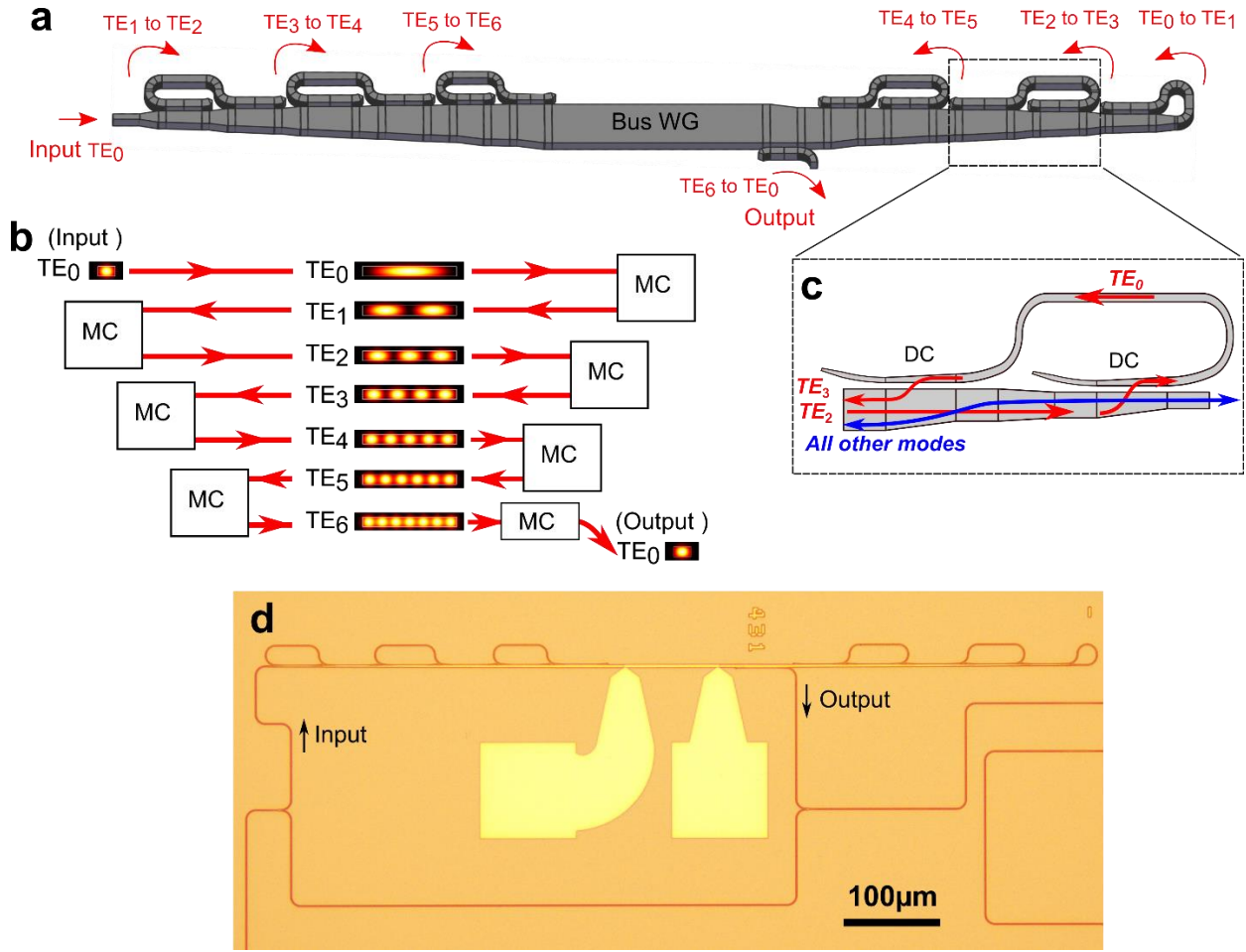

**Figure 1 | Multipass recycling structure based on mode multiplexing.** (**a**) Schematic (not to scale) of a 7-pass structure that utilizes seven spatial modes for recycling. (**b**) Schematic description of the light path, showing mode conversion at each pass. (**c**) Schematic (not to scale) of a structure that converts the $TE_2$ mode to the $TE_3$ mode and reverses the propagation direction, while transmitting all other lower-order modes. (**d**) Optical microscope image of a 7-pass recycling structure embedded in a MZI. A resistive heater is fabricated on top of the $SiO_2$ cladding to induce phase shifting via the thermo-optic effect. MC: Mode converter. DC: Directional coupler. WG: Waveguide.

Using the recycling platform we demonstrate the ability to enhance the power efficiency of a thermo-optic phase shifter by more than 8 times and achieve a low power consumption of $P_\pi =$ 1.7 mW for a $\pi$ phase shift. By embedding a thermo-optic phase shifter in the recycling structure, light passes the phase shifter multiple times and accumulates the phase shift from all passes. In order to measure the accumulated phase shift, we embed the recycling structure in a Mach-Zehnder interferometer (MZI), as shown by Fig. 1d. We measure the phase shift of the MZI interference fringes at different applied powers. For comparison, we perform the same measurement on devices with 3-pass, 5-pass, and 7-pass recycling structures, as well as a standard 1-pass device without recycling. All these devices have the same bus waveguide width of 2.44 $\mu$m under the thermo-optic phase shifter. We observe a clear enhancement of the power efficiency as the number of recycling passes increases, as shown in Fig. 2. We measure $P_\pi$ of 15.4 mW, 4.6 mW, 2.6 mW, and 1.7 mW for the 1-pass, 3-pass, 5-pass, and 7-pass phase shifters, respectively. This corresponds to a power-efficiency enhancement of 3.3, 5.9, and 8.9 times in the 3-pass, 5-pass, and 7-pass phase shifter, respectively. Note that the factor of enhancement is slightly higher than the number of passes. This is because the effective refractive indices of the higher-order modes are more sensitive to temperature change due to stronger dispersion (see Supplementary Note 1). We measure a thermal time constant $\tau$ of 6.5 $\mu$s, independent of the number of passes (see Supplementary Fig. 7). The power-time product $P_\pi \cdot \tau$ figure of merit of our 7-pass recycling-enhanced phase shifter is 11.1 mW·$\mu$s. It is smaller than other state-of-art thermo-optic phase shifters[2-4], including heaters on thermally-isolated free-standing waveguides[4] ($P_\pi \cdot \tau$ =56.1 mW·$\mu$s) and doped-silicon heaters with adiabatic bends[2] ($P_\pi \cdot \tau$ =30.5 mW·$\mu$s).

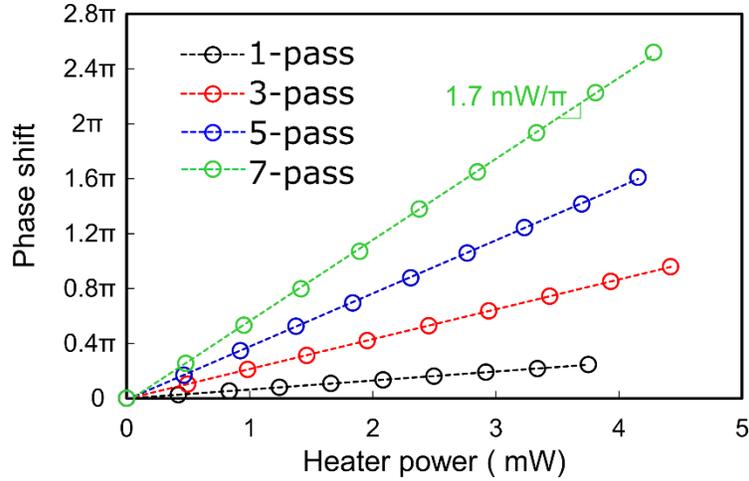

**Figure 2 | Measured phase shift in the recycling-enhanced phase shifters.** The accumulated phase shift induced by the 3-pass, 5-pass and 7-pass recycling-enhanced thermo-optic phase shifters near the wavelength of 1600 nm, extracted by measuring the transmission spectra of the MZIs. As a reference we also show a phase shifter with no recycling (denoted by 1-pass). The dashed lines are the linear fits to the data. One can see an enhancement of phase shifter efficiency by 3.3, 5.9 and 8.9 times with the 3-pass, 5-pass and 7-pass recycling structure, respectively.

We show light recycling with a bandwidth exceeding 100 nm and a low insertion loss of 0.44 dB per recycling event for the first 5 passes. In Fig. 3 we show the measured insertion losses of the recycling structures extracted from the visibility of the MZI interference fringes. One can see that the minimal insertion losses are 1.2 dB (at the wavelength of 1570 nm), 2.2 dB (at 1594 nm), and 4.6 dB (at 1601 nm) for the 3-pass, 5-pass, and 7-pass recycling structure, respectively. The 3-dB bandwidths of all recycling structures exceed 100 nm. From these measured results, we estimate that the first 5 passes have an insertion loss of 0.44 dB per recycling event. The sixth and seventh recycling have higher insertion losses of 1.2 dB per recycling event. Note that these insertion losses are not fundamental and are due to the fabrication-induced variation.

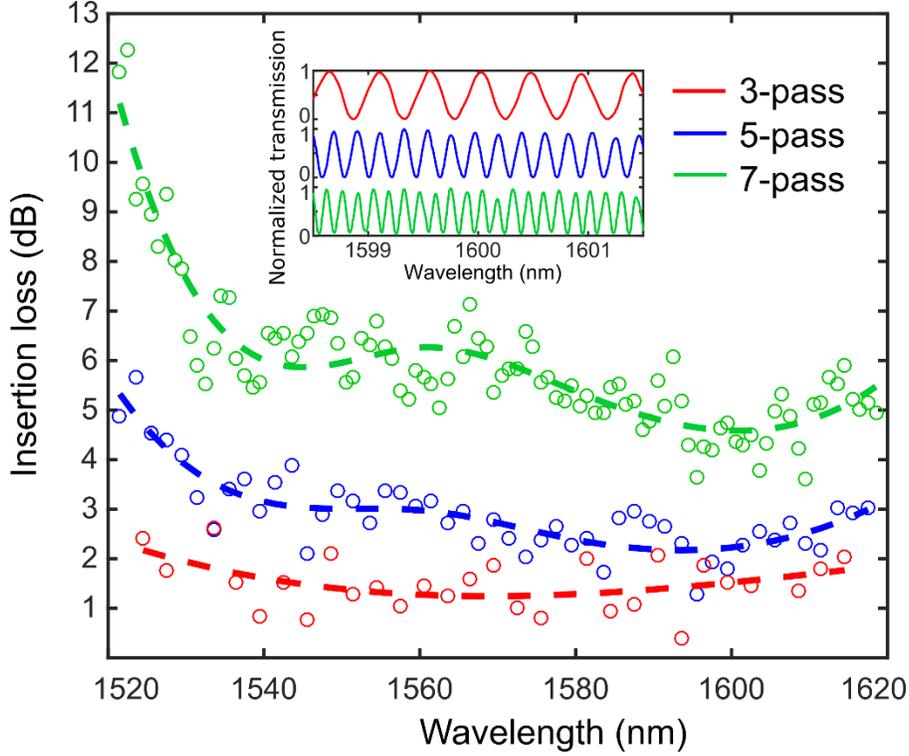

**Figure 3 | Measured insertion losses of the 3-pass, 5-pass and 7-pass recycling structures (circles) extracted from the visibility of the MZI interference fringes.** The dashed lines are cubic spline fits that exclude the artifacts due to the interference from facet reflections. One can see that the 3-dB bandwidths, in which the insertion losses remain less than 3 dB above the minimums, are at least 100 nm for all structures. The inset shows the measured transmission spectra of the MZIs with different recycling structures. One can see that the free spectral range of the interference fringes decreases as the number of passes increases, demonstrating an increase of the optical path length due to light recycling.

To minimize the effect of fabrication on the variations of the optimal dimensions relative to the wavelength, we design the platform to operate for a wide range of wavelengths by ensuring that no interference effect is used. In order to convert each of the spatial modes, we use adiabatic directional couplers[23-26] that in contrast to traditional interference-based directional couplers[19,21-22], do not require precise phase matching and exact length. In adiabatic couplers, the local effective refractive index $n_{eff}$ of the modes is tuned adiabatically by tapering the widths of the two coupled waveguides. By tapering each of the waveguides in opposite directions, we ensure that the modes of the two individual waveguides have the same $n_{eff}$ somewhere along the center of the coupler. These two individual modes couple and form supermodes ($S_1$ and $S_2$). In contrast to a traditional interference-based directional coupler, in which both supermodes are excited and interfere with each other, in an adiabatic directional coupler only one of the supermodes is excited. As an example we show in Fig. 4 an adiabatic directional coupler that couples the $TE_3$ mode of the bus waveguide ($TE_3^{bus}$) to the $TE_0$ mode of the access waveguide ($TE_0^{access}$). Figure 4a shows the local effective refractive index $n_{eff}(z)$ at each cross section along the directional coupler. One can see that the $n_{eff}$ of the two individual modes ($TE_3^{bus}$ and $TE_0^{access}$) cross each other at the center of the directional coupler, resulting in an anticrossing between the supermode $S_1$ and $S_2$. Fig. 4b shows the schematic of this adiabatic directional coupler. Fig. 4c shows the mode profile of the super mode $S_1$, which evolves from the $TE_3^{bus}$ mode at the input to the $TE_0^{access}$ mode at the output. We simulate the coupling efficiency of each adiabatic directional coupler, from which we calculate the insertion losses of the recycling structures in the presence of dimensional variation, as shown in Fig. 5a. The simulated insertion losses remain less than 0.6 dB, 1.5 dB, and 2.8 dB for the 3-pass, 5-pass and 7-pass recycling, respectively, if the dimensional variation is controlled within ± 15 nm. Our simulation shows that the insertion loss of the 7-pass recycling can potentially be reduced to only 2 dB if the fabrication variations in all the geometries are kept within ± 11 nm. This large tolerance of our platform is in stark contrast to the required subnanometer tolerance in resonators used often for recycling. Our platform requires no active compensation since this tolerance is practically achievable in both electron beam and deep ultraviolet lithography[15]. We show in Fig. 5b the simulated insertion losses of the recycling structures at different wavelengths. All structures show theoretical 3-dB bandwidths exceeding 200 nm. These large fabrication tolerance and high bandwidth are enabled by our design of adiabatic directional couplers and cannot be achieved with traditional interference-based directional couplers (see Supplementary Note 3 for a detailed comparison between the two types of couplers).

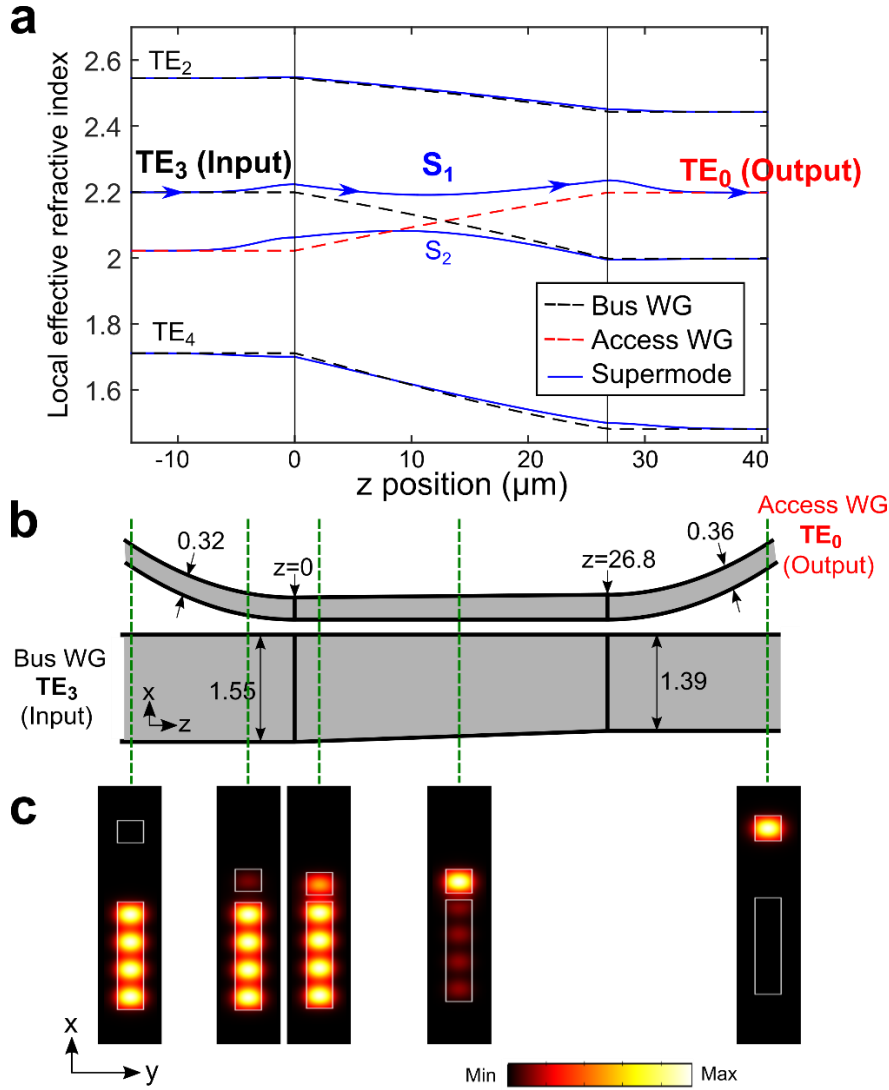

**Figure 4 | Design of the mode-selective adiabatic directional coupler.** (**a**) The local effective refractive indices $n_{eff}(z)$ at each cross section of the $TE_3^{bus}$-to-$TE_0^{access}$ adiabatic directional coupler. One can see that the $n_{eff}(z)$ of the two individual waveguide (in the absence of the other waveguide) (black and red dashed lines) cross at the center of the directional coupler. The $n_{eff}(z)$ of the supermodes of the coupled waveguide system is also shown (blue). (**b**) Schematic of the $TE_3^{bus}$-to-$TE_0^{access}$ adiabatic directional coupler. The access waveguide and the bus waveguide are tapered linearly in opposite directions. (**c**) Mode profile of the supermode $S_1$ at different cross

sections along the *z* position (indicated by the green dashed lines in **b**). The colour represents the energy density of the mode. The cross sections of the waveguides are indicated by white lines.

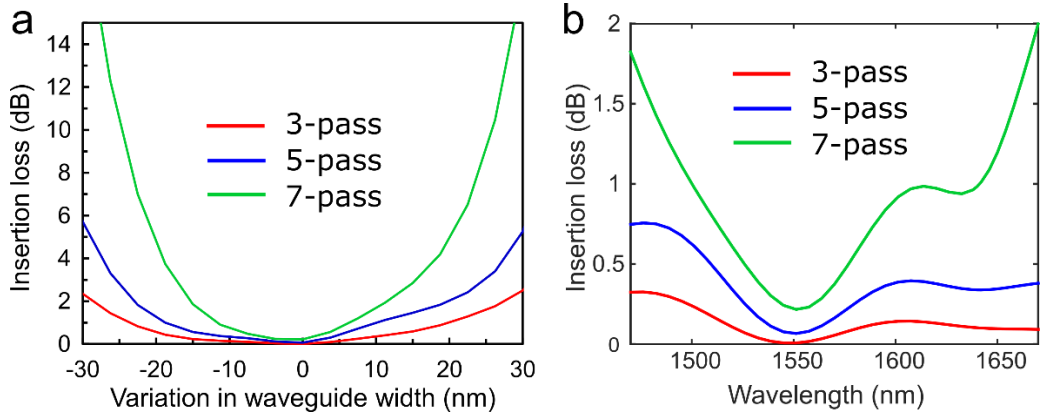

**Figure 5 | Simulated tolerance to fabrication-induced geometrical changes and bandwidth of the 3-pass, 5-pass and 7-pass recycling structures.** (**a**) Simulated insertion losses as functions of the variation in the access waveguide width, the dimension that induces the largest insertion loss variation. The insertion losses remain less than 2.8 dB for a variation of $\pm$ 15 nm, indicating good tolerance to fabrication variation. (**b**) Simulated insertion losses as functions of wavelength, assuming no dimensional variation. It shows that the 3-dB bandwidths of all recycling structures exceed 200 nm.

In conclusion, we demonstrate a recycling approach that could enable silicon photonics to overcome one of its main hurdles: the lack of scalability in power. This light recycling approach provides fabrication tolerance and bandwidth that are fundamentally unachievable in resonators traditionally used for recycling, which for enabling *F* round trips of recycling, are limited to only $1/F$ of the free spectral range in bandwidth. This limited bandwidth is a fundamental consequence of the time-bandwidth limitation[27], which cannot be overcome by lowering the quality factor of resonators. Our approach can be applied to enhance the efficiency of phase shifters in a large-scale optical phased array[12-14]. It can also be used to enhance signal strength for broadband applications such as spectroscopy, where resonator-based light recycling has only limited applications due to the need of a extended bandwidth[28,29]. Although we use thermo-optic phase shifters as a demonstration, our approach can enhance the performance of a variety of photonic devices, such as electro-optic phase shifters, electro-absorption modulators, and delay lines. The adiabatic directional couplers can be further optimized with novel techniques that ensures adiabaticity in a smaller footprint[24,30].

## Methods

**Device fabrication.** We fabricate the devices on a silicon-on-insulator (SOI) wafer from Soitec, which has a 250 nm top silicon layer and a 3 $\mu$m buried oxide layer. The silicon wire waveguides are patterned by electron beam lithography (Elionix ELS-G100) using ma-N 2403 photoresist, etched by inductively coupled plasma (ICP) using fluorine-based chemistry, and clad with 1 $\mu$m silicon dioxide ($SiO_2$) by plasma-enhanced chemical vapor deposition (PECVD). On top of the $SiO_2$ cladding, we fabricate 100 $\mu$m-long, 3 $\mu$m-wide resistive heaters as thermo-optic phase shifters. The heaters and the pads are patterned with electron beam lithography, followed by sputtering (5 nm Ti and 100 nm W) and lift-off.

**Device simulation.** We simulate the effective refractive indices of the waveguides with the finite-element method using the commercial package COMSOL Multiphysics 5.1. The local effective refractive indices shown in Fig. 4a are simulated with the waveguide cross section at each z position along the directional coupler. We simulate the coupling efficiencies of the directional couplers with the commercial package Fimmwave 6.4. We also use Fimmwave to ensure that the tapers and the bends of the adiabatic directional couplers vary slowly enough to satisfy the adiabatic condition. We confirm that almost all the power is in one supermode as light propagates along the directional coupler, and the mode profile of this supermode evolves completely from one waveguide to another (as in Fig. 4c). We calculate the insertion loss of the multipass recycling structure from the product of the coupling efficiencies of all individual directional couplers. The detailed dimensions of the devices we use in the simulation are given in Supplementary Table 1. Unless stated otherwise, we perform the simulations at the wavelength of 1550 nm.

**Device characterization.** We measure the transmission spectra of the devices with a tunable laser source (Ando AQ4321D) that can sweep the wavelength from 1520 nm to 1620 nm. We couple the light into the chip and collect it from the chip with lensed fibres. The silicon waveguide is inversely tapered to a 160 nm width at the input and output facets to match the mode size of the lensed fibres. We use 1×2 multimode interference (MMI) couplers with 50/50 split ratio to split and recombine light in the MZI. The output power is detected by a power meter with an InGaAs detector (Newport 818-IG) when acquiring the device transmission spectra. Another InGaAs detector with a higher speed (Thorlabs PDB150C) is used when measuring the temporal response of the resistive heaters. The resistive heaters are driven with a source meter (Keithley 2400-LV) and a function generator in the phase shift and heater temporal response measurements, respectively. The resistance of the resistive heaters (excluding the contact resistance) range from 835 Ω to 983 Ω in our devices. We calculate applied heater power from the measured current (with Keithley 2400-LV source meter) and from the resistance of each heater.

**Extraction of the insertion loss from the visibility of interference fringes.** The insertion loss of the recycling structure is directly related to the visibility of the MZI interference fringes, assuming the splitters of the MZI have an accurate split ratio of 50/50. The visibility $V$ of interference fringes is defined as $(P_{max} - P_{min})/(P_{max} + P_{min})$, where $P_{max}$ and $P_{min}$ are the power at the peak and the valley, respectively. The insertion loss in dB is given by $-log_{10}\{2[1 - \sqrt{(1 - V^2)}]/V^2 - 1\}$. We fit the acquired fringes locally with sine functions, extract the visibility, and calculate the insertion loss as a function of wavelength.

## Acknowledgements

The authors gratefully acknowledge support from MOABB DARPA program contracted by Trex, inc under award # HR0011-16-C-0107, from ONR for award # N00014-16-1-2219, and NSF Engineering Research Center for Integrated Access Networks (CIAN). This work was performed in part at the Advanced Science Research Center at the City College of New York.

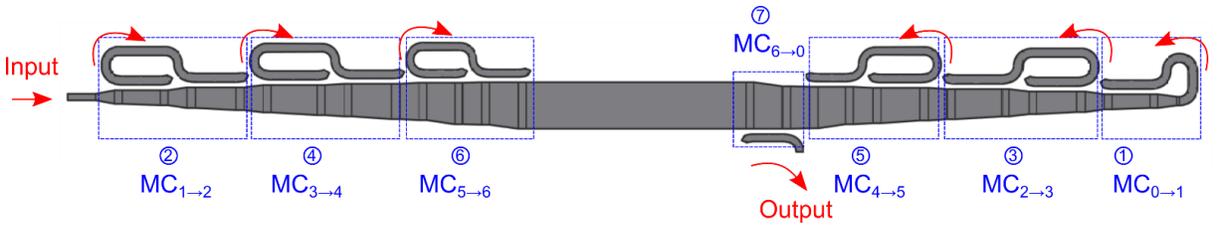

**Supplementary Figure 1 | Schematic (not to scale) of a 7-pass recycling structure that utilizes the seven modes from the TE$_0$ to the TE$_6$ mode.** The labels of ① to ⑦ represent of the order of recycling sequence. MC$_{i \rightarrow j}$ represents the mode converter that converts the TE$_i$ mode to the TE$_j$ mode.

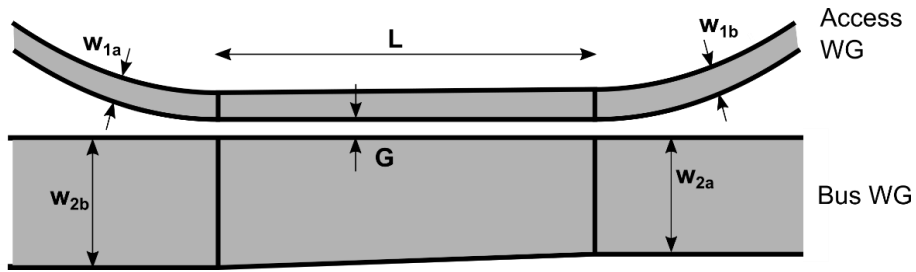

**Supplementary Figure 2 | Schematic of a directional coupler.** The dimensions are given in Supplementary Table 1 and 2.

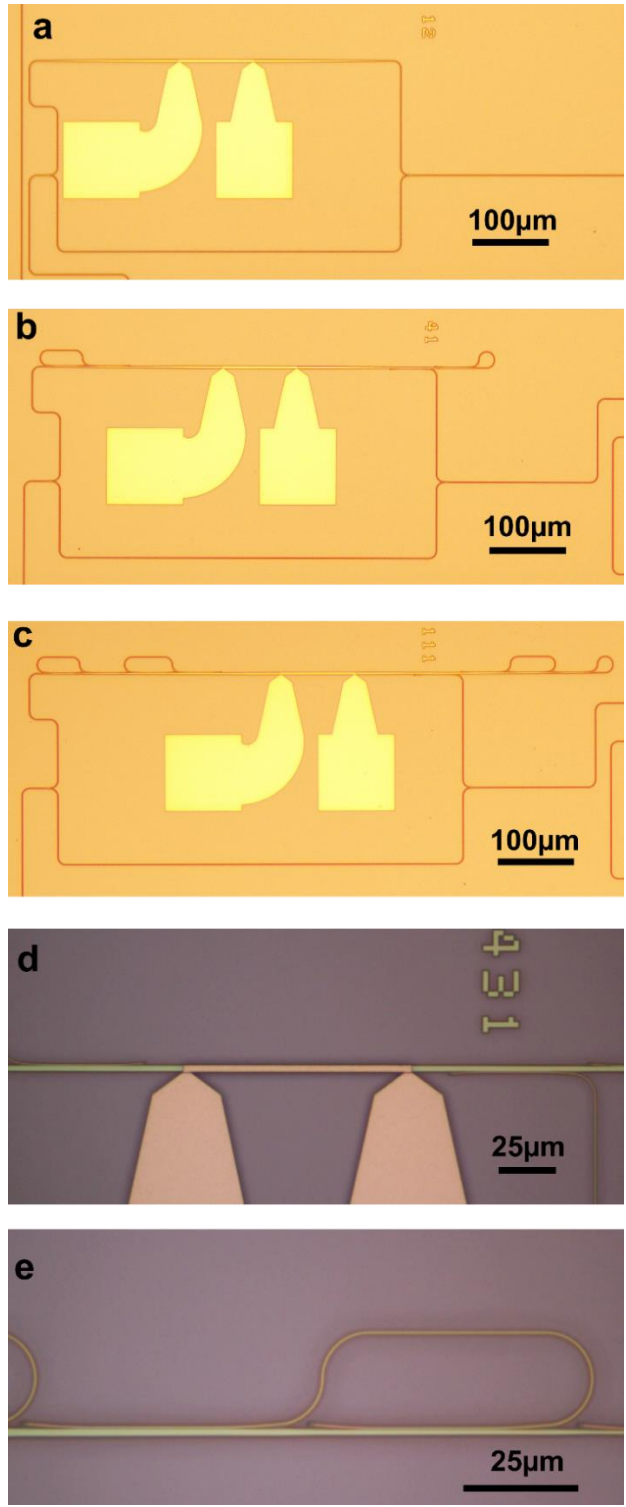

**Supplementary Figure 3| Optical microscope images of the Mach-Zehnder interferometers with different recycling structures.** (**a**) The standard MZI without recycling structure. (**b**) The MZI with the 3-pass recycling structure. (**c**) The MZI with the 5-pass recycling structure. (**d**) The resistive heater in the 7-pass recycling structure. (**e**) The $TE_2$-to-$TE_3$ mode converter in the 7-pass recycling structure.

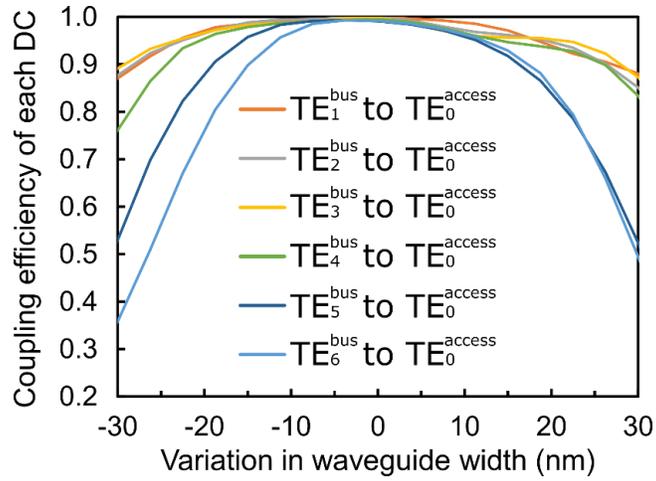

**Supplementary Figure 4 | Simulated dimensional tolerance of different mode-selective adiabatic directional couplers.** In this simulation, we give the same variation to the access waveguide widths $w_{1a}$ and $w_{1b}$, which are the most crucial dimensions. The coupling efficiencies remains > 90% within a variation of ± 15 nm, indicating good tolerance to fabrication variation. The higher-order mode directional couplers are less tolerant due to smaller $n_{eff}$ difference between modes. The notations $w_{1a}$ and $w_{1b}$ are defined in Supplementary Fig. 2.

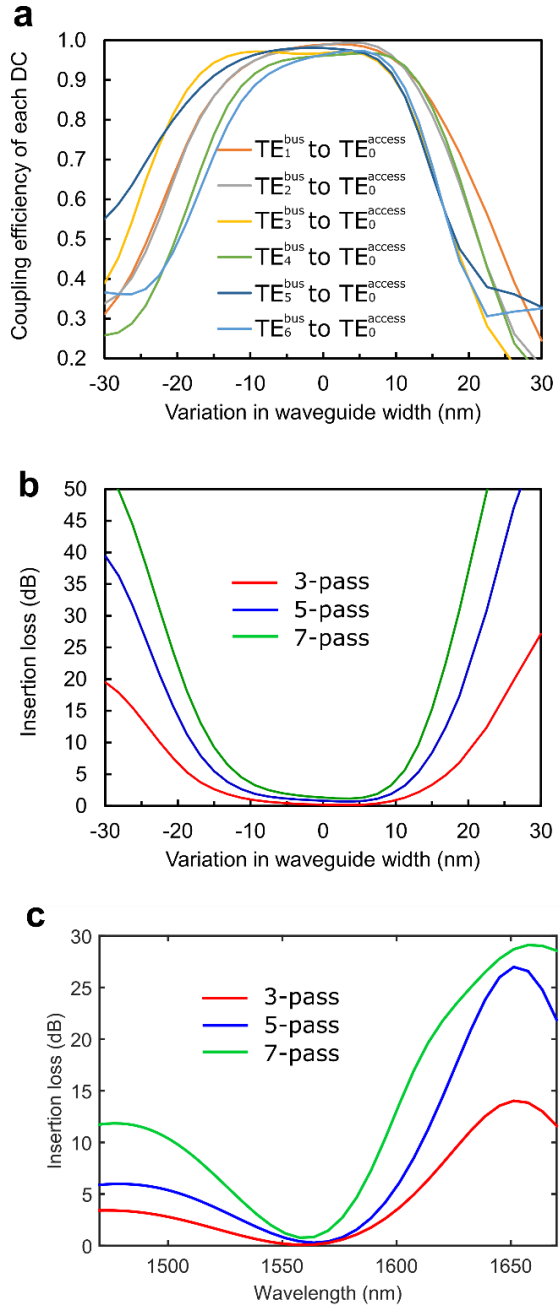

**Supplementary Figure 5 | Simulated performance of interference-based directional couplers.** (**a**) Simulated coupling efficiencies of each interference-based directional couplers (as opposed to the adiabatic couplers we use) as functions of the variation in the access waveguide width. The detailed dimensions of these directional couplers are given in Supplementary Table 2. (**b**) Simulated tolerance of the recycling structures that use the interference-based directional couplers. (**c**) Simulated insertion loss spectra of the recycling structures that use the interference-based directional couplers. We perform this simulation to draw a comparison between the adiabatic directional couplers and the traditional interference-based directional couplers. The simulation shows that the interference-based directional couplers are less tolerant to fabrication variation compared to the adiabatic counterpart. The coupling efficiency can decrease to 66 % when there is a dimensional variation of $\pm$ 15 nm. The 3-dB optical bandwidths of the 3-pass, 5-pass and 7-pass recycling structures that use the interference-based directional couplers are 110 nm, 65 nm and 45nm, respectively, which are much narrower than the adiabatic design.

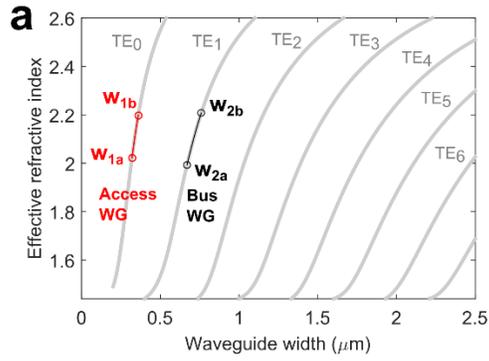
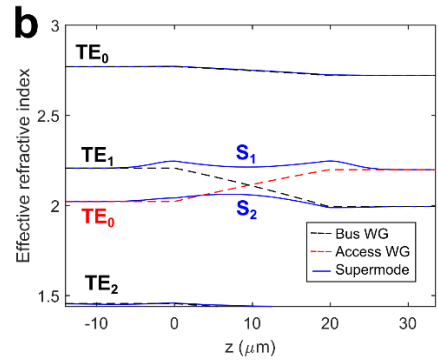
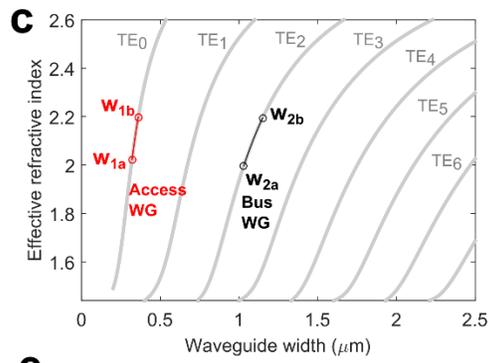
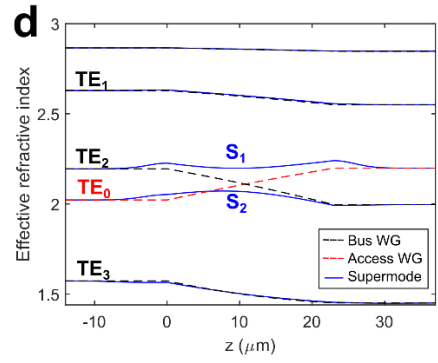
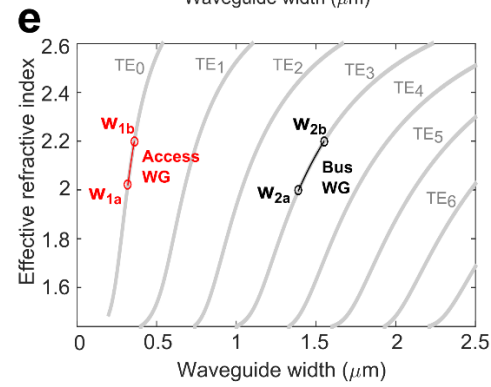
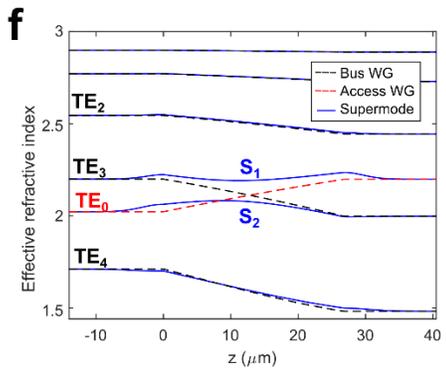

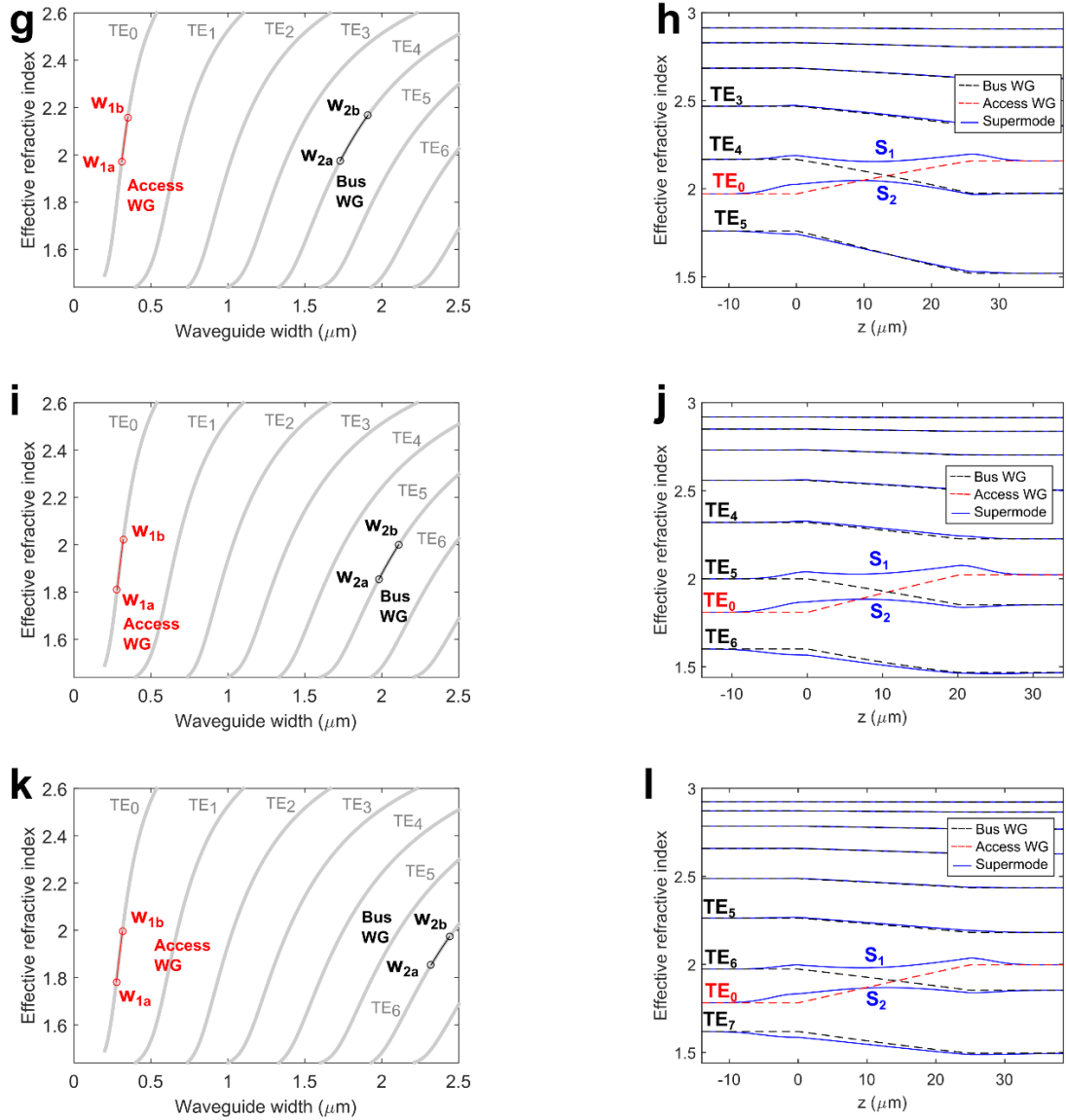

**Supplementary Figure 6 | Local effective refractive indices of the adiabatic directional couplers designed for different modes.** (**a,c,e,g,i,k**) The simulated $n_{eff}$ of different waveguide modes as functions of the waveguide width, showing the ability to engineer the $n_{eff}$ by controlling the waveguide width[1-3]. The widths used in the $TE_1^{bus}$-to-$TE_0^{access}$, $TE_2^{bus}$-to-$TE_0^{access}$, $TE_3^{bus}$-to-$TE_0^{access}$, $TE_4^{bus}$-to-$TE_0^{access}$, $TE_5^{bus}$-to-$TE_0^{access}$, $TE_6^{bus}$-to-$TE_0^{access}$ adiabatic directional coupler are marked in (a), (c), (e), (g), (i), (k), respectively. We take advantage of the high index contrast in silicon waveguides, which leads to the large separation between the $n_{eff}$ of different modes. (**b,d,f,h,j,l**) The simulated local effective refractive indices $n_{eff}(z)$ at each cross section in the adiabatic directional couplers. The $n_{eff}(z)$ of the $TE_1^{bus}$-to-$TE_0^{access}$, $TE_2^{bus}$-to-$TE_0^{access}$, $TE_3^{bus}$-to-$TE_0^{access}$, $TE_4^{bus}$-to-$TE_0^{access}$, $TE_5^{bus}$-to-$TE_0^{access}$, $TE_6^{bus}$-to-$TE_0^{access}$ adiabatic directional couplers are shown in (b), (d), (f), (h), (j), (l), respectively. The $n_{eff}(z)$ of the two individual waveguides (in the absence of the other waveguide) is plotted as black and red dashed lines. We achieve mode selectivity by allowing only one crossing between the red and black dashed lines. The $n_{eff}(z)$ of the supermodes of the coupled waveguide system is plotted as blue solid lines. The notations of waveguide widths are defined in Supplementary Fig. 2. The detailed dimensions of the adiabatic directional couplers are given in Supplementary Table 1.

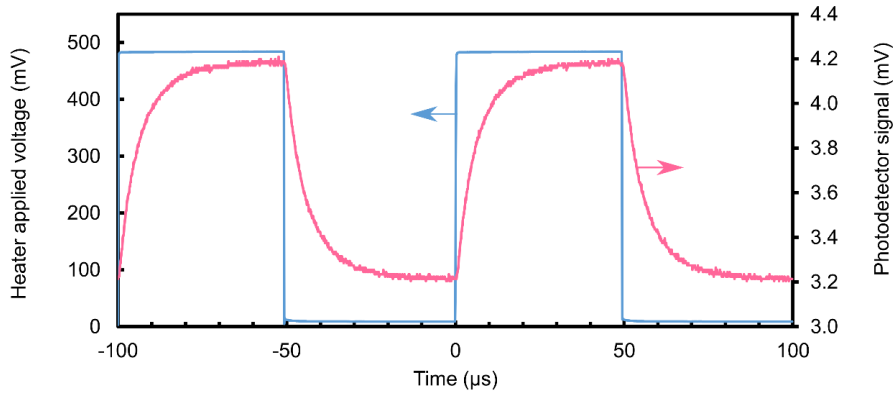

**Supplementary Figure 7 | The temporal response of the thermo-optic phase shifter with the 7-pass recycling structure.** The thermal time constant is obtained by fitting the photodetector signal with an exponential function. We find that the thermal time constant is independent of the number of recycling. In all 1-pass, 3-pass, 5-pass and 7-pass devices, we measure thermal time constants of $6.4 \pm 0.2$ $\mu$s (rise time) and $6.6 \pm 0.4$ $\mu$s (fall time).

**Supplementary Table 1 | The dimensions of the adiabatic directional couplers used in the recycling structure.** The radius of the arc is 130 $\mu$m. The height of the silicon waveguide is 250 nm. Other notations are defined in Supplementary Fig. 2.

| Mode in the access waveguide | Mode in the bus waveguide | $w_{1a}$ ($\mu$m) | $w_{1b}$ ($\mu$m) | $w_{2a}$ ($\mu$m) | $w_{2b}$ ($\mu$m) | G ($\mu$m) | L ($\mu$m) |
|---|---|---|---|---|---|---|---|
| $TE_0$ | $TE_1$ | 0.32 | 0.36 | 0.67 | 0.76 | 0.1 | 19.8 |
| $TE_0$ | $TE_2$ | 0.32 | 0.36 | 1.03 | 1.15 | 0.1 | 23 |
| $TE_0$ | $TE_3$ | 0.32 | 0.36 | 1.39 | 1.55 | 0.1 | 26.8 |
| $TE_0$ | $TE_4$ | 0.31 | 0.35 | 1.73 | 1.91 | 0.1 | 25.8 |
| $TE_0$ | $TE_5$ | 0.28 | 0.32 | 1.98 | 2.11 | 0.1 | 20.2 |
| $TE_0$ | $TE_6$ | 0.275 | 0.315 | 2.32 | 2.44 | 0.13 | 24.9 |

**Supplementary Table 2 | The dimensions of an alternative design that uses interference-based directional couplers.** We design these directional couplers to draw a comparison between the adiabatic and the traditional interference-based directional couplers. The bus waveguide is slightly tapered to relax the requirement of exact phase matching[4], but the coupling is not in the adiabatic regime. The radius of the arc is 40 $\mu$m. The height of the silicon waveguide is 250 nm. Other notations are defined in Supplementary Fig. 2. The simulated performance of these directional couplers is shown in Supplementary Fig. 5.

| Mode in the access waveguide | Mode in the bus waveguide | $w_{1a}$ ($\mu$m) | $w_{1b}$ ($\mu$m) | $w_{2a}$ ($\mu$m) | $w_{2b}$ ($\mu$m) | G ($\mu$m) | L ($\mu$m) |
|---|---|---|---|---|---|---|---|
| $TE_0$ | $TE_1$ | 0.39 | 0.39 | 0.77 | 0.85 | 0.1 | 17.4 |
| $TE_0$ | $TE_2$ | 0.39 | 0.39 | 1.18 | 1.3 | 0.1 | 21.6 |
| $TE_0$ | $TE_3$ | 0.39 | 0.39 | 1.58 | 1.74 | 0.1 | 23 |
| $TE_0$ | $TE_4$ | 0.36 | 0.36 | 1.89 | 2.05 | 0.1 | 19.8 |
| $TE_0$ | $TE_5$ | 0.35 | 0.35 | 2.15 | 2.43 | 0.1 | 30.1 |
| $TE_0$ | $TE_6$ | 0.33 | 0.33 | 2.47 | 2.71 | 0.1 | 30.1 |

**Supplementary Note 1 | Theoretical enhancement factor of the recycling-enhanced phase shifters.** In the recycling-enhanced phase shifters, the phase shift is enhanced because the light passes through the same heated waveguide multiple times. The total phase shift is the sum of the phase shift seen by the light in each pass. However, because in each pass the light passes with a different mode, it sees a slightly different phase shift. We use the finite-element method to simulate $dn_{eff,m}/dT$ for all modes. Here $n_{eff,m}$ denotes the effective refractive index of the TE$_{m-1}$ mode, and $T$ is the temperature. We perform the simulation at the wavelength of 1600 nm. The thermo-optic coefficients of Si and SiO$_2$ used in the simulation are $1.86\times10^{-4}$ and $9.5\times10^{-6}$ respectively[5,6]. For a silicon waveguide with a cross section of 2.44 µm × 0.25 µm, we obtain $dn_{eff,m}/dT$ of $1.883\times10^{-4}$, $1.916\times10^{-4}$, $1.976\times10^{-4}$, $2.069\times10^{-4}$, $2.207\times10^{-4}$, $2.407\times10^{-4}$, and $2.641\times10^{-4}$ for the TE$_0$ to TE$_6$ mode, respectively. One can see that the higher-order modes have slightly larger $dn_{eff,m}/dT$ due to stronger dispersion. The theoretical enhancement factor for an *n*-pass phase shifter is given by

$$\left[\sum_{m=1}^{n}\left(\frac{dn_{eff,m}}{dT}\right)\right]/\frac{dn_{eff,1}}{dT}. \quad (1)$$

We obtain the theoretical enhancement factor for the 3-pass, 5-pass and 7-pass phase shifters, which are 3.07, 5.34 and 8.02 respectively. Note that this analysis considers only the phase shift created in the waveguide section that is directly under the resistive heater. Other nearby waveguide sections also have a small temperature increase and therefore contribute to the phase shift. We attribute the small deviation between the measured and theoretical enhancement factors to such contribution.

**Supplementary Note 2 | Free spectral range of the Mach-Zehnder interferometers with recycling structures.** We confirm light recycling by measuring the free spectral range (FSR) of the Mach-Zehnder interferometers (MZIs) with recycling structures. Because light circulates multiple times in the recycling structure, the optical path length of the recycling arm becomes longer than the physical length. The FSR of a MZI is given by $\lambda^2/\Delta OPL$, where $\Delta OPL$ is the optical path difference between the two interferometer arms[7]. Because of recycling, the MZI has a larger $\Delta OPL$ and therefore a smaller FSR compared to a MZI without recycling. We measure the transmission spectra of the MZIs with different number of passes, as shown in the inset of Fig. 3 of the main text. The measured FSR of the MZI interference fringes near the wavelength 1600 nm are 0.461 nm, 0.214 nm, and 0.141 nm for the 3-pass, 5-pass, and 7-pass MZIs, respectively. The measured FSR decreases with the number of recycling passes, indicating an increase of $\Delta OPL$.

To confirm the recycling mechanism, we calculate the theoretical FSR of each MZI from the optical path length of our design and compare with the experimental results. The optical path length is given by $\int n_g(l)dl$, where $n_g$ is the group index. The group index depends on the dimensions and the mode of a waveguide. We use the finite-element method to simulate the group index of the relevant modes of the designed waveguides. We integrate the group index along the path that light recycles. The theoretical FSR near the wavelength of 1600 nm are 0.472 nm, 0.206 nm, and 0.145 nm for the 3-pass, 5-pass, and 7-pass MZIs, respectively. The good agreement between the measured and theoretical FSR confirms the mechanism of mode conversion and light circulation. We attribute the small deviation (less than 4 %) to the non-perfect dimension in actual fabrication.

**Supplementary Note 3 | Comparison between adiabatic directional couplers and interference-based directional couplers.** For comparison, we simulate an alternative design of recycling structures that uses interference-based directional couplers instead of adiabatic directional couplers. In Supplementary Table 2 we show the dimensions of the interference-based directional couplers we use in the simulation. In Supplementary Fig. 5 we show the performance of this alternative design. In stark contrast to the adiabatic design, the 7-pass recycling structure using traditional interference-based directional couplers has a high insertion loss of 15.3 dB with the same dimensional variation of ± 15 nm and a 3-dB bandwidth of only 45 nm.

**Supplementary References**